\documentclass[a4paper,11pt]{article}
\usepackage{amsfonts, amsmath, amssymb}
\usepackage[utf8]{inputenc}
\usepackage[a4paper,top=3cm,bottom=2.5cm,left=2.5cm,right=2.5cm, bindingoffset=5mm]{geometry}
\usepackage{color}
\usepackage{booktabs}
\usepackage{subfig}
\usepackage{graphicx}
\usepackage{float}
\usepackage{hyperref}
\hypersetup{citecolor=blue}
\usepackage{array}
\usepackage{colordvi}
\usepackage{multirow}
\usepackage{threeparttable}

\begin{document}

\vspace{20pt}
\begin{center}

{\bf{\Large An $S_4$ model inspired from self-complementary neutrino mixing}}

\vspace{0.4cm}  Xinyi Zhang$\mbox{}^{1,2}$ \footnote{E-mail: \texttt{xzhang\_phy@pku.edu.cn}}

\vspace{0.1cm}
$^1${\em INPAC, SKLPPC, and Department of Physics,}

{\em Shanghai Jiao Tong University, 200240 Shanghai, China.}

$^2${\em School of Physics, Peking University, 100871 Beijing, China.}

\end{center}

\begin{abstract}
We build an $S_4$ model for neutrino masses and mixings based on the self-complementary (SC) neutrino mixing pattern. The SC mixing is constructed from the self-complementarity relation plus $\delta_{\rm CP}=-\frac{\pi}{2}$. We elaborately construct the model at a percent level of accuracy to reproduce the structure given by the SC mixing. After performing a numerical study on the model's parameter space, we find that in the case of normal ordering, the model can give predictions on the observables that are compatible with their $3\sigma$ ranges, and give predictions for the not-yet observed quantities like the lightest neutrino mass $m_1\in [0.003,0.010]$ eV and the Dirac CP violating phase $\delta_{\rm CP}\in[256.72^\circ,283.33^\circ]$.
\end{abstract}

\section{Introduction}

Neutrinos are known to be massive and have mixing. Among the various ways to understand neutrino masses and mixing, discrete flavour symmetries are widely used in describing the mixing structure as well as offering testable predictions~\cite{King:2015aea,Petcov:2014laa}. $S_4$ is one of the popular groups that are adopted as symmetries of the flavour sector in the literature. It has been used to reproduce the tribimaximal (TB) mixing~\cite{Lam:2008rs,Bazzocchi:2008ej,Ding:2009iy,Ishimori:2010xk,Zhao:2011pv}, bimaximal (BM) mixing~\cite{Altarelli:2009gn,Toorop:2010yh,Meloni:2011fx} and trimaximal (TM) mixing~\cite{King:2011zj,Krishnan:2012me,Varzielas:2012pa,Luhn:2013vna}, and is combined with CP symmetry to give realistic mixing predictions~\cite{Ding:2013hpa,Feruglio:2013hia,Li:2014eia}. Among many discrete groups used for model building, $S_4$ exceeds in being small and having three-dimensional irreducible representations. Compared to the group $A_4$ which is smaller and also have a three-dimensional irreducible representation, it is argued that $S_4$ is more suitable to reproduce BM mixing at the leading order~\cite{Altarelli:2009gn}.

In the meantime, phenomenological observations can give us hints of underlying structure, and are used as starting points for further model building. The above mentioned constant mixing patterns inspire many discrete flavour symmetry models. 

The self-complementarity relation (SC) of lepton mixing~\cite{Zhang:2012xu,Zheng:2011uz} is observed in 2012 in light of the relatively large value of $\theta_{13}$ measured by reactor neutrino experiments~\cite{An:2012eh,Ahn:2012nd}. It correlates the three lepton mixing angles in a simple way. Namely, the sum of the two relative small mixing angles are equal to the large one, and to $45^\circ$~\footnote{In ref~\cite{Zhang:2012xu}, this is the self-complementarity relation of the first kind. There are two other kinds of self-complementarity relations, which are the sum of the two smaller angles equal to the third angle, or to $45^\circ$. }. There are several phenomenological investigations related to this relation~\cite{Zhang:2012xu,Zhang:2012pv,Qu:2013aca,Ke:2014hxa}, but a model realization of this relation is still missing. The current work presents as a primary trial in this situation. By ``primary" we intentionally keep the model ``small" by leaving the extension to quark sector, to supersymmetry and/or GUTs to further studies.

In this work, we firstly construct a self-complementary mixing pattern from the self-complementarity relation and $\delta_{\rm CP}=-\frac{\pi}{2}$. Then we build a neutrino mass model with $S_4$ flavour symmetry, whose breaking renders the mass matrix structure given by the self-complementary mixing. We perform a numerical study on the model's parameter space and discuss the model's predictions for neutrino masses and mixing in the end.

This paper is organized as follows. In Section~\ref{sec:sc}, we introduce the self-complementary mixing, and give the structure of the Majorana neutrino mass matrix dictated by the self-complementary mixing. The explicit model building is in Section~\ref{sec:model}. Then we perform a numerical analysis in Section~\ref{sec:ph} for the phenomenology of the model. In Section~\ref{sec:sum} we make the summary and give the conclusion.

\section{General approach and preparations}\label{sec:sc}

In this section we first construct a mixing pattern featuring the self-complementarity relation, and then use this mixing pattern to get the neutrino mass matrix. In next section we will reproduce the neutrino mass matrix  structure in the context of a model. It is in the same spirit of many models in the literatures. 


\subsection{Self-complementary mixing}
We start with the self-complementarity relation of the first kind~\cite{Zhang:2012xu}, i.e.,
\begin{align}
\theta_{12}+\theta_{13}=\theta_{23}=45^\circ,
\end{align}
where $\theta_{ij}$ are lepton mixing angles in the standard parametrization. Compared with the latest global fit result in Table~\ref{tab:glb}, we see that both relations work marginally at the $3\sigma$ edge. Nevertheless, there are several possibilities to account for the deviation from exact SC relation: the SC relation may hold at high energy scale (e.g. a seesaw scale), so it receives corrections from renormalization group (RG) running effects; it may receive corrections from high order operators; it may hold only in the neutrino sector, and receive corrections from the charged lepton sector. A quantitative examination is model dependent and we will see this soon in the case of our model. So we conclude that the comparison with the current data does not weaken the necessity of the present work.

\begin{table}
	\centering
	\begin{threeparttable}
\caption{Global fit result for the mixing parameters~\cite{deSalas:2017kay}}\label{tab:glb}
\begin{tabular}{l|c|c}
\toprule\midrule
 & bfp$\pm 1\sigma$ & $3\sigma$ range\\
 \hline
 $\theta_{12}~~[^\circ]$&$34.5^{+1.1}_{-1.0}$&$31.5-38.0$\\
 $\theta_{23}~~[^\circ]$&$41.0\pm 1.1$~(N)\tnote{*},~$50.5\pm 1.0$~(I)&$38.3-52.8$~(N),~$38.5-53.0$~(I)\\
 $\theta_{13}~~[^\circ]$&$8.44^{+0.18}_{-0.15}$~(N),~$8.41^{+0.16}_{-0.17}$~(I)&$7.9-8.9$\\
 $\delta_{\rm CP}~[^\circ]$&$252^{+56}_{-36}$~(N),~$259^{+47}_{-41}$~(I)&$0-360$~(N),~$142-360$~(I)\\
 \bottomrule 
\end{tabular}
\begin{tablenotes}
	\item[*]N (I) stands for normal (inverted) ordering.
\end{tablenotes}
\end{threeparttable}
\end{table}

An additional ingredient we use for constructing the self-complementary mixing is $\delta_{\rm CP}=-\frac{\pi}{2}$. We use it for two reasons: firstly, it is the value of $\delta_{\rm CP}$ which is indicated by T2K~\cite{Abe:2013hdq} and NO$\nu$A~\cite{Bian:2015opa}, and is within the $1\sigma$ range of the global fit~\ref{tab:glb}; secondly, it is special in the sense that $\delta_{\rm CP}$ contributes its maximal to the Jarlskog invariant. Applying the SC relation together with $\delta_{\rm CP}=-\frac{\pi}{2}$ to the standard parameterization, we get the self-complementary mixing directly as
\begin{align}
U_{\rm SC}=\left(
\begin{array}{ccc}
 \cos \left(\frac{\pi }{4}-\theta_{13}\right) \cos \theta_{13} &  \sin \left(\frac{\pi }{4}-\theta_{13}\right) \cos \theta_{13}& i \sin \theta_{13} \\
 \frac{i\cos \left(\frac{\pi }{4}-\theta_{13}\right) \sin\theta_{13}-\sin \left(\frac{\pi }{4}-\theta_{13}\right)}{\sqrt{2}} & \frac{\cos \left(\frac{\pi }{4}-\theta_{13}\right)+i \sin\left(\frac{\pi }{4}-\theta_{13}\right) \sin \theta_{13}}{\sqrt{2}}& \frac{\cos \theta_{13}}{\sqrt{2}} \\
 \frac{\sin \left(\frac{\pi }{4}-\theta_{13}\right)+i \cos \left(\frac{\pi }{4}-\theta_{13}\right) \sin \theta_{13}}{\sqrt{2}} & \frac{i \sin \left(\frac{\pi }{4}-\theta_{13}\right) \sin \theta_{13}-\cos \left(\frac{\pi }{4}-\theta_{13}\right)}{\sqrt{2}} & \frac{\cos\theta_{13}}{\sqrt{2}} \\
\end{array}
\right),\label{eq:Usc}
\end{align}
and the whole lepton mixing matrix when neutrinos are Majorana particles is $U_{\rm PMNS}=U_{\rm SC}.P,~~P={\rm Diag}\{e^{-i\alpha_1/2},e^{-i\alpha_2/2},1\}$~\cite{Bilenky:1980cx}.

Here we comment on why we have to construct a mass matrix perturbatively to realize a self-complementary mixing. As can be seen from Eq.~(\ref{eq:Usc}), the SC mixing satisfies $|U_{\rm SC}|_{{\rm \mu} i}=|U_{\rm SC}|_{\tau i}$, which is the $\mu$-$\tau$ exchange symmetry prediction for the mixing~\cite{Zhang:2014rua}. The immediate guess would be if we try to construct the mass matrix using Eq.~(\ref{eq:Usc}) directly, we will arrive at a mass matrix resembling the $\mu$-$\tau$ exchange symmetry. As $|U_{\rm SC}|_{\mu i}=|U_{\rm SC}|_{\tau i}$ is necessary but not sufficient to get $\mu$-$\tau$ exchange symmetry, the guess is wrong. However, the realistic situation is similar: a mass matrix given by Eq.~(\ref{eq:Usc}), which is at the meantime simple enough for model building, only reflects the two input: $\theta_{23}=45^\circ,~\delta_{\rm CP}=-\frac{\pi}{2}$. That is to say, the other ingredient of SC mixing, i.e., $\theta_{12}+\theta_{13}=45^\circ$ is obscured from the direct construction of a mass matrix. This ingredient gives substructure of a mass matrix that is given by $\theta_{23}=45^\circ,~\delta_{\rm CP}=-\frac{\pi}{2}$. To see this substructure, we have to construct the mass matrix perturbatively. 

There is a constant mixing pattern having the features including the SC relation and also a maximal CP violating phase~\cite{Qu:2013aca}. For the reason stated above, we do not use it for SC model building here. There are $A_4$ models featuring $\theta_{23}=45^\circ,~\delta_{\rm CP}=-\frac{\pi}{2}$~\cite{He:2015afa}.

\subsection{Mass matrix structure inspired from SC mixing}
By identifying
\begin{align}
\sin\theta_{13}&=\lambda,\\
\cos\theta_{13}&\cong 1-\frac{1}{2}\lambda^2,
\end{align}
we get the expansion of $U_{\rm SC}$ in powers of $\lambda$,
\begin{align}
U_{\rm SC}&\equiv U_{\lambda^0}+\lambda U_{\lambda^1}+\lambda^2 U_{\lambda^2}+...\nonumber\\
&= \left(
\begin{array}{ccc}
\frac{1}{\sqrt{2}} & \frac{1}{\sqrt{2}} & 0 \\
-\frac{1}{2} &\frac{1}{2} &\frac{1}{\sqrt{2}} \\
\frac{1}{2} &-\frac{1}{2} & \frac{1}{\sqrt{2}}\\
\end{array}\right) +\lambda\left(
\begin{array}{ccc}
\frac{1}{\sqrt{2}} & -\frac{1}{\sqrt{2}} & i \\
\frac{1}{2}+\frac{i}{2} &\frac{1}{2}+\frac{i}{2} &0 \\
-\frac{1}{2}+\frac{i}{2} &-\frac{1}{2}+\frac{i}{2} & 0\\
\end{array}\right)+\lambda^2\left(
\begin{array}{ccc}
-\frac{1}{\sqrt{2}} & -\frac{1}{\sqrt{2}} & 0 \\
\frac{1}{4}+\frac{i}{2} &-\frac{1}{4}-\frac{i}{2} &-\frac{1}{2\sqrt{2}} \\
-\frac{1}{4}+\frac{i}{2} &\frac{1}{4}-\frac{i}{2} & -\frac{1}{2\sqrt{2}}\\
\end{array}\right)~~~\nonumber\\
&+...~ .~~~\label{eq:Usc_exp}
\end{align}

We use the expansion of  $ U_{\rm SC}$ to get the Majorana mass matrix expanded in $\lambda$,
\begin{align}
\hat{m}_\nu&= U_{\rm SC}^* \hat{m}^d U_{\rm SC}^\dagger\nonumber\\
&=( U_{\lambda^0}+\lambda U_{\lambda^1}+\lambda^2 U_{\lambda^2}+...)^* \hat{m}^d ( U_{\lambda^0}+\lambda U_{\lambda^1}+\lambda^2 U_{\lambda^2}+...)^\dagger\nonumber\\
&\equiv \hat{m}_0 + \lambda \hat{m}_1 + \lambda^2 \hat{m}_2 +...,
\end{align}
where $\hat{m}^d=diag\{m_1,m_2,m_3\}$. We use a hat notation above $m$ to distinguish matrix $\hat{m}_i$ from eigenvalues $m_i$. The mass matrix of the first few orders of $\lambda$ reads
\begin{align}
\hat{m}_0&= U_{\lambda^0}^* \hat{m}^d U_{\lambda^0}^\dagger;\\
\hat{m}_1&= U_{\lambda^0}^* \hat{m}^d U_{\lambda^1}^\dagger+U_{\lambda^1}^* \hat{m}^d U_{\lambda^0}^\dagger;\\
\hat{m}_2&= U_{\lambda^0}^* \hat{m}^d U_{\lambda^2}^\dagger+U_{\lambda^1}^* \hat{m}^d U_{\lambda^1}^\dagger+U_{\lambda^2}^* \hat{m}^d U_{\lambda^0}^\dagger;\\
...\nonumber
\end{align}

Using the $U_{\lambda^i}$ in Eq.~(\ref{eq:Usc_exp}), the mass matrix can be constructed accordingly. 

At the leading order, we get
\begin{align}
\hat{m}_0 =\left(
\begin{array}{ccc}
 \frac{1}{2}(m_1+m_2) & \frac{1}{2 \sqrt{2}}(m_2-m_1) & \frac{1}{2 \sqrt{2}} (m_1-m_2)\\
 \frac{1}{2 \sqrt{2}} (m_2-m_1)& \frac{1}{4} (m_1+m_2+2 m_3) & \frac{1}{4} (-m_1-m_2+2 m_3) \\
 \frac{1}{2 \sqrt{2}}(m_1-m_2) & \frac{1}{4} (-m_1-m_2+2 m_3) & \frac{1}{4} (m_1+m_2+2 m_3) \\
\end{array}
\right),
\end{align}
which is of the form
\begin{align}
\hat{m}_0\sim \left(
\begin{array}{ccc}
x & y & -y \\
y & z & z-x \\
-y & z-x & z\\
\end{array}\right),\label{eq:m0}
\end{align}
where x, y and z are in general complex and their complexity comes solely from the Majorana phases which we associated with the eigenvalues $m_i$ in $\hat{m}^d$. Such a mass matrix can be diagonalized by 
\begin{align}
U_0=\left(
\begin{array}{ccc}
\frac{1}{\sqrt{2}} & -\frac{1}{\sqrt{2}} & 0 \\
-\frac{1}{2} & -\frac{1}{2} &\frac{1}{\sqrt{2}} \\
\frac{1}{2} &\frac{1}{2} & \frac{1}{\sqrt{2}}\\
\end{array}\right),
\end{align}
which is the BM mixing~\cite{Bimaximal} \footnote{A diagonal matrix $P={\rm Diag}\{1,-1,1\}$ needs to be multiplied on the left of $U_0$ to make it the same as the BM mixing.}. Actually, this is the main reason for our choice of the flavour group to be $S_4$.

At $\mathcal{O}(\lambda)$, we find $\hat{m}_1$ is of the following form
\begin{align}
\hat{m}_1= a\left(
\begin{array}{ccc}
2 & 0 & 0 \\
0 & -1+i & 1 \\
0 & 1 & -1-i\\
\end{array}\right)+
b\left(
\begin{array}{ccc}
0 & i & i \\
i & 0 & 0 \\
i & 0 & 0\\
\end{array}\right),\label{eq:m1}\\
a=\frac{1}{2}(m_1-m_2),\quad b=-\frac{1}{2\sqrt{2}}(m_1+m_2+2m_3),~\label{eq:ab}
\end{align}
where a, b are complex numbers.

At $\mathcal{O}(\lambda^2)$, we find $\hat{m}_2$ is of the following form
\begin{align}
\hat{m}_2= c\left(
\begin{array}{ccc}
2 & 0 & 0 \\
0 & 1 & 1 \\
0 & 1 & 1\\
\end{array}\right)+
d\left(
\begin{array}{ccc}
0 & 1 & -1 \\
1 & 0 & 0 \\
-1 & 0 & 0\\
\end{array}\right)+
e\left(
\begin{array}{ccc}
0 & i & i \\
i & 0 & 0 \\
i & 0 & 0\\
\end{array}\right),\label{eq:m2}\\
c=-\frac{1}{4}(m_1+m_2+2m_3),\quad d=\frac{5}{4\sqrt{2}}(m_1-m_2),\quad e=-\frac{1}{\sqrt{2}}(m_1-m_2)~\label{eq:cde},
\end{align}
where c, d and e are complex, too.

Since $\lambda=\sin\theta_{13}\simeq0.15$, we will stop at $\mathcal{O}(\lambda^2)$, so the deviation from the exact SC mixing at percent level is expected. In next section, we will build a model to reproduce the neutrino mass matrix structure to this order.
 
\section{Model construction}\label{sec:model}
In this section we build a model to reproduce the neutrino mass matrix structure inspired from the SC mixing. As a first trial to do this, we keep the model non-supersymmetric, and leave the discussion of the quark sector to future works. We adopt here an $S_4$ flavour symmetry. The standard model lepton doublets are assigned to 3 representation of $S_4$. We extend the standard model by adding three right handed neutrinos, which explain the smallness of light neutrino masses through seesaw type I~\cite{seesawI}. The right handed neutrinos are gauge singlets, but in 3 representation of $S_4$. Through the spontaneous breaking of the $S_4$ symmetry, which occurred when the flavons acquire their vacuum expectation values (vev), we get the structure of the Majorana mass matrix that will result in the desired neutrino mass matrix structure after applying seesaw. 

We list first the field representations in $S_4$ and charges under additional symmetries in our model in Table~\ref{tab:assignment}. Higgs is the singlet in $S_4$ and is not charged under any of the additional symmetries, so it is omitted in the table. We have flavons in 1, 2 and 3 representations of $S_4$. The $\rm U(1)$ charges are arranged in a way that no new terms at the discussed order will show up, explicitly, $x\neq m\neq n\neq z$. 

\begin{table}
\centering\caption{Field representations in $S_4$ and charges under additional symmetries of the model}\label{tab:assignment}
\resizebox*{1.1\textwidth}{.14\textheight}{%
\begin{tabular}{l|ccccccccccccccccccc}
\toprule\midrule
& L & $e_R$ & $\mu_R$ & $\tau_R$ & N & $\phi_e$ & $\phi_\mu$ & $\phi_\tau$ & $\theta$ & $\xi_1$ & $\phi_1$ & $\psi_1$ & $\phi_{21}$ & $\phi_{22}$ & $\phi_{23}$ & $\phi_{31}$ & $\phi_{32}$ & $\psi_3$ & $\xi_3$\\
\hline
$S_4$ & 3 & 1 & 1 & 1 & 3 & 3 & 3 & 3 & 1 & 1 & 3 & 2 & 3 & 3 & 3 & 3 & 3 & 2 & 1\\
$\rm U(1)$ & -x & z & m & n & x & $\frac{1}{2}(x-z)$ & x-m & x-n & 0 &-2x & -2x & -2x & -x & -x & -x & $-\frac{2}{3}x$ & $-\frac{2}{3}x$ & y & -2x-2y\\
$\rm U(1)_{\rm FN}$    & 0 & 2 & 1 & 0 & 0 & 0 & 0& 0 & -1 & 0 & 0 & 0 & 0 & 0 & 0 & 0 & 0 & 0 & 0\\
$\mathbb{Z}_2$ & 0 & 0  & 0  & 0 & 0 & 0 & 0& 0 & 0 & 0 & 0 & 0 & 1 & 0 & 0 & 0 & 0 & 0 & 0\\
$\mathbb{Z}_2$ & 0 & 0  & 0  & 0 & 0 & 0 & 0& 0 & 0 & 0 & 0 & 0 & 0 & 1 & 0 & 0 & 0 & 0 & 0\\
$\mathbb{Z}_2$ & 0 & 0  & 0  & 0 & 0 & 0 & 0& 0 & 0 & 0 & 0 & 0 & 0 & 0 & 1 & 0 & 0 & 0 & 0\\
$\mathbb{Z}_3$ & 0 & 0  & 0  & 0 & 0 & 0 & 0& 0 & 0 & 0 & 0 & 0 & 0 & 0 & 0 & 1 & 2 & 0 & 0\\
\bottomrule
\end{tabular}}%
\end{table}

The effective Lagrangian reads
\begin{align}
-\mathcal{L}&= y_e\bar{L}\tilde{H}e_{\rm R} \left(\frac{\phi_e}{\Lambda}\right)^2\left(\frac{\theta}{\Lambda}\right)^2
+y_\mu\bar{L}\tilde{H}\mu_{\rm R} \left(\frac{\phi_\mu}{\Lambda}\right)\left(\frac{\theta}{\Lambda}\right)
+y_\tau\bar{L}\tilde{H}\tau_{\rm R} \left(\frac{\phi_\tau}{\Lambda}\right)
+y_\nu\bar{L}HN\nonumber\\
&+\frac{y_{11}}{\Lambda}(NN)_1\xi_1+\frac{y_{12}}{\Lambda}(NN)_2\psi_1+\frac{y_{13}}{\Lambda}(NN)_3\phi_1\nonumber\\
&+\frac{y_{21}}{\Lambda^2}(N\phi_{21})_3(N\phi_{21})_3
+\frac{y_{22}}{\Lambda^2}(NN)_3(\phi_{22}\phi_{22})_3
+\frac{y_{23}}{\Lambda^2}(NN)_3(\phi_{23}\phi_{23})_3\nonumber\\
&+\frac{y_{31}}{\Lambda^3}(N\psi_3)_3(N\psi_3)_3\xi_3
+\frac{y_{32}}{\Lambda^3}\left((NN)_3\phi_{31}\right)_1(\phi_{31}\phi_{31})_1
+\frac{y_{33}}{\Lambda^3}\left((NN)_3\phi_{32}\right)_1(\phi_{32}\phi_{32})_1,~~~~~~~
\end{align}
where the flavons coupling to the charged leptons are marked with flavour indices while flavons coupling to neutrinos are marked with numbers 1,2,3; besides, ``$\phi$" stands for a triplet flavon in $S_4$, ``$\psi$" for a doublet  and ``$\xi$" for a singlet. In the neutrino sector, we denote the $S_4$ contraction with a subscript outside the parenthesis. Since there is no ambiguity in the $S_4$ contraction in the charged lepton sector, the contraction is not explicitly written. The couplings $y_{ij}$ in the Majorana neutrino mass term are of mass dimension $1$ and are at the scale of heavy neutrino mass (we use a notation ``y" instead of ``m" to avoid confusion with various m in this model); the $\Lambda$ denotes the cutoff scale of the theory.

In such a model setup, the structure of the neutrino mass matrix of interest comes solely from $S_4$ breaking flavon vevs. The additional symmetries are used to forbid the unwanted terms from the Lagrangian. The $\mathbb{Z}_n$ symmetries are needed to distinguish the copies of flavons in the same $S_4$ representation, e.g., forbiding terms like $NN\phi_{21}\phi_{22}$. The $\rm U(1)$ symmetry forbid terms like $\bar{L}\tilde{H}e_R\phi_\mu\theta^2$. The $\rm U(1)_{\rm FN}$ symmetry is responsible for the hierarchical masses of charged leptons. The potential Goldstone boson coming from the spontaneous breaking of $\rm U(1)$ symmetry may be gauged away by adding more particles, which is beyond the scope of the current work. It is also possible to use more cyclic symmetries instead of the $\rm U(1)$ symmetry to complete the same construction. 
\subsection{The charged lepton sector}

In the charged lepton sector, we adopt the same idea as in Ref.~\cite{Altarelli:2009gn} that a $\rm U(1)_{\rm FN}$ symmetry~\cite{Froggatt:1978nt} is used in combination with the other additional symmetries, to generate the hierarchical masses of the charged leptons. When the flavons acquire the following vevs\footnote{An illustrative example of how we get the flavon vev structure is shown in Appendix~\ref{sec:vev}}, 
\begin{align}
\langle\phi_e\rangle=v_{\phi_e}(0,0,1),\quad\langle\phi_\mu\rangle=v_{\phi_\mu}(0,0,1),\quad\langle\phi_\tau\rangle=v_{\phi_\tau}(0,1,0),
\end{align}
the charged lepton mass matrix reads
\begin{align}
\hat{m}_l=\left(\begin{array}{ccc}
    \frac{y_e}{\Lambda^4}v v_{\theta}^2 v_{\phi_e}^2 & 0 & 0\\
    0 & \frac{y_\mu}{\Lambda^2}v v_\theta v_{\phi_\mu} & 0\\
    0 & 0 & \frac{y_\tau}{\Lambda}v v_{\phi_\tau}\\
    \end{array}\right),
\end{align}
which exhibits a residual $\mathbb{Z}_4$ symmetry of $S_4$.

As a rough estimation, we require that all the charged lepton Yukawa coupling constants are of the same maginitude: $y_e\sim y_\mu \sim y_\tau$; and all the flavons are of the same magnitude: $v_{\phi_e} \sim v_{\phi_\mu} \equiv v_{\phi_l}$. Comparing with $\frac{m_e}{m_\mu}|_{\rm expt}\simeq 0.005,\quad \frac{m_\mu}{m_\tau}|_{\rm expt}\simeq 0.06$, we get $\frac{v_{\phi_l}}{\Lambda}\sim 0.08, \quad \frac{v_\theta}{\Lambda}\sim 0.06$, which is the same as in Ref.~\cite{Altarelli:2009gn}.

For the field assignments as in Table.~\ref{tab:assignment}, the charged lepton sector and the neutrino sector are well separated. In the meantime, since the right handed charged leptons are all singlet of $S_4$, the resulting charged lepton mass matrix will always be diagonal (even when higher order operators enter). As a result, there will be no corrections to the lepton mixing matrix from the charged lepton sector. So we will focus on the neutrino part of the model from now on.

\subsection{The neutrino sector}

The Dirac neutrino mass matrix coming from the neutrino Yukawa coupling reads
\begin{align}
\hat{m}_{\rm D}=y_\nu v\left(\begin{array}{ccc}
1 & 0 & 0\\
0 & 0 & 1\\
0 & 1 & 0\\
\end{array}\right).
\end{align}

We construct the Majorana neutrino mass matrix order by order. In the leading order, the flavons vevs are
\begin{align}
\langle\xi_1\rangle=v_{\xi_1},\quad\langle\phi_1\rangle=v_{\phi_1}(0,1,1),\quad\langle\psi_1\rangle=v_{\psi_1}(1,\frac{1}{3}).
\end{align}
The resulting Majorana neutrino mass matrix when the flavons acquire their vevs is
\begin{align}
\hat{m}_{\rm LO}=\frac{y_{11}}{\Lambda}v_{\xi_1}\left(\begin{array}{ccc}
1& 0  & 0\\
0  & 0 & 1\\
0 & 1& 0\\
\end{array}\right)+\frac{y_{12}}{\Lambda}v_{\psi_1}\left(\begin{array}{ccc}
1& 0  & 0\\
0  & \frac{\sqrt{3}}{6} & -\frac{1}{2}\\
0 & -\frac{1}{2}& \frac{\sqrt{3}}{6}\\
\end{array}\right)\frac{y_{13}}{\Lambda}v_{\phi_1}\left(\begin{array}{ccc}
0& -1  &1\\
-1  & 0 & 0\\
1 & 0& 0\\
\end{array}\right),
\end{align}
which resembles the structure of $\hat{m}_0$ in Eq.~(\ref{eq:m0}). The singlet flavon $\xi_1$ is necessary to avoid $m_3=0$ when making direct comparison with $\hat{m}_0$. 

In the next-to-leading order, the flavon vevs are
\begin{align}
\langle\phi_{21}\rangle=v_{\phi_{21}}(0,1,1),\quad \langle\phi_{22}\rangle=v_{\phi_{22}}(0,1,0),\quad\langle\phi_{23}\rangle=v_{\phi_{23}}(1,1,1),
\end{align}
the resulting mass matrix is
\begin{align}
\hat{m}_{\rm NLO}&=\frac{y_{21}}{\Lambda^2}v_{\phi_{21}}^2
    \left(\begin{array}{ccc}
        -2 & 0 & 0\\
        0  & 1 &-1\\
        0  & -1& 1\\
        \end{array}\right)  
    +\frac{y_{22}}{\Lambda^2}v_{\phi_{22}}^2  
       \left(\begin{array}{ccc}
            0  & 0 & 0\\
            0  & 1 & 0\\
            0  & 0& -1\\
            \end{array}\right) \nonumber\\
    &+\frac{y_{23}}{\Lambda^2}v_{\phi_{23}}^2  
        \left(\begin{array}{ccc}
                0  & -2 & -2\\
                -2  & 0 & 0\\
                -2  & 0 & 0\\
                \end{array}\right).                       
\end{align}
Here we use two terms to reproduce the ``a" term in Eq.~(\ref{eq:m1}). Since we adopt a real representation of $S_4$ as in Ref.~\cite{Altarelli:2009gn}, we arrange the imaginary unit to the coefficient of the constant matrix. We do the same to the ``b" term and other terms whenever necessary.

In the next-to-next leading order, the flavon vevs are
\begin{align}
\langle\xi_3\rangle=v_{\xi_3},\quad\langle\psi_3\rangle=v_{\psi_3}(-\sqrt{3},1),\quad\langle\phi_{31}\rangle=v_{\phi_{31}}(0,1,1),\quad\langle\phi_{32}\rangle=v_{\phi_{32}}(0,-1,1).
\end{align}
The resulting mass matrix is
\begin{align}
\hat{m}_{\rm NNLO}=\frac{y_{31}}{\Lambda^3}v_{\psi_3}^2v_{\xi_3}
    \left(\begin{array}{ccc}
        3 & 0 & 0\\
        0 & \frac{3}{2} & \frac{3}{2}\\
        0 & \frac{3}{2} & \frac{3}{2}\\
        \end{array}\right)
    +\frac{y_{32}}{\Lambda^3}v_{\phi_{31}}^3
        \left(\begin{array}{ccc}
            0 & -2 & 2\\
            -2 & 0 & 0 \\
            2 & 0 & 0\\
            \end{array}\right) 
    +\frac{y_{33}}{\Lambda^3}v_{\phi_{32}}^3
           \left(\begin{array}{ccc}
               0 & -2 & -2\\
               -2 & 0 & 0 \\
               -2 & 0 & 0\\
               \end{array}\right),                    
\end{align}
which resembles the structure of $\hat{m}_2$ in Eq.~(\ref{eq:m2}).

Due to the strict constraints given by all the symmetries of the model, we find no higher order terms contributing to the Majorana neutrino mass matrix to mass dimension $10$. 

The effective light neutrino mass matrix is given by seesaw as
\begin{align}
\hat{m}_{\nu_{\rm model}}=\hat{m}_{\rm D}(\hat{m}_{\rm LO}+\hat{m}_{\rm NLO}+\hat{m}_{\rm NNLO}+...)^{-1}\hat{m}_{\rm D}^T\equiv \hat{m}_{\rm D}\hat{m}_{\rm R}^{-1}\hat{m}_{\rm D}^T,\label{eq:mmodel}
\end{align}
where we define $\hat{m}_{\rm R}=\hat{m}_{\rm LO}+\hat{m}_{\rm NLO}+\hat{m}_{\rm NNLO}+...$ as the heavy neutrino mass matrix. It is expected from the above construction that $\hat{m}_{\rm R}$ resembles the same structure inspired from the SC mixing to order $\lambda^2$.

The parameters in $\hat{m}_{\rm R}$ can be simplified by noticing that the coefficients ``a,b,c,d,e" are all of the form ``$m_1+m_2+2m_3$" or ``$m_1-m_2$". By forcing $\hat{m}_{\nu_{\rm model}}$ in the same form as the one constructed from the SC mixing to $\mathcal{O}(\lambda^2)$ (which also means in the same form as $\hat{m}_{\rm R}$), we arrive at five independent real parameters in $\hat{m}_{\rm R}$: $\tilde{v}_{\xi_1}$, $\tilde{v}_{\psi_1}$ and their phase $\gamma$; $\tilde{v}_{\phi_1}$ and its phase $\rho$. For a detailed description of the parameters simplification, see Appendix~\ref{apd:para}.

Although our approach here is in the same spirit of the usual model building of this kind, there is something different: we do not require that $\hat{m}_{\nu_{\rm model}}$ is diagonalized by the SC mixing. We presume $\hat{m}_{\nu_{\rm model}}$ resembles the structure of $\hat{m}_{\rm R}$. However, since $\hat{m}_{\rm D}$ does not commute with $\hat{m}_{\rm R}$, $\hat{m}_{\nu_{\rm model}}$ cannot be of exactly the same structure as $\hat{m}_{\rm R}$. This fact motivates a detailed scan of the parameter space to check the validity of our model, as we will do in next section.

\section{Phenomenology}\label{sec:ph}

\indent In this section we perform a numerical scan of the parameter space of the model. The seesaw scale is set at $10^{13}$ GeV, and $y_\nu$ is fixed to 0.1. We scan over five real parameters: $\tilde{v}_{\xi_1}$, $\tilde{v}_{\psi_1}$ and their phase $\gamma$; $\tilde{v}_{\phi_1}$ and its phase $\rho$. Since the model is constructed at a high energy, we use the REAP package~\cite{Antusch:2005gp} to perform the evolution of the mixing parameters from the seesaw scale to the low energy scale to make comparison with the oscillation observables. 

\begin{figure}
\centering
\includegraphics[scale=0.49]{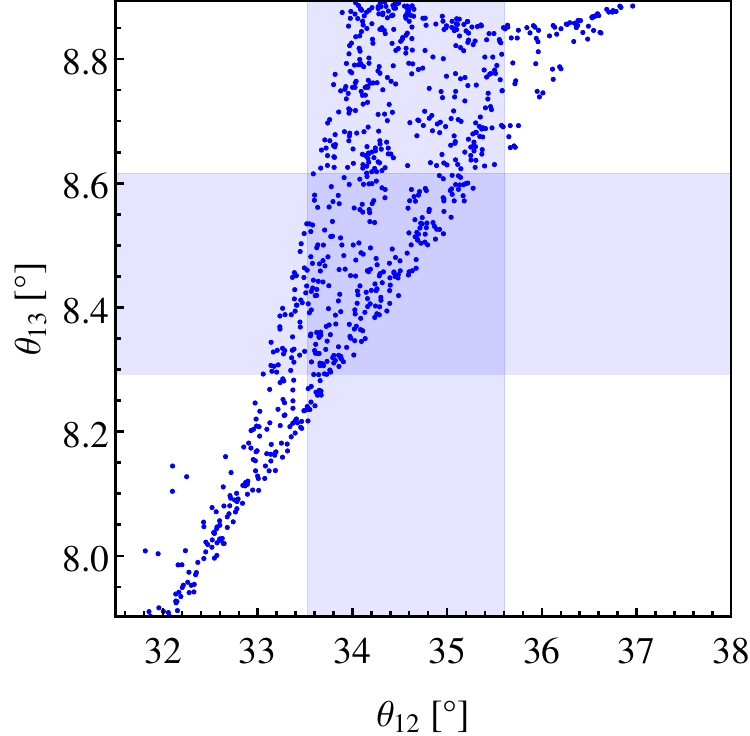}
\includegraphics[scale=0.49]{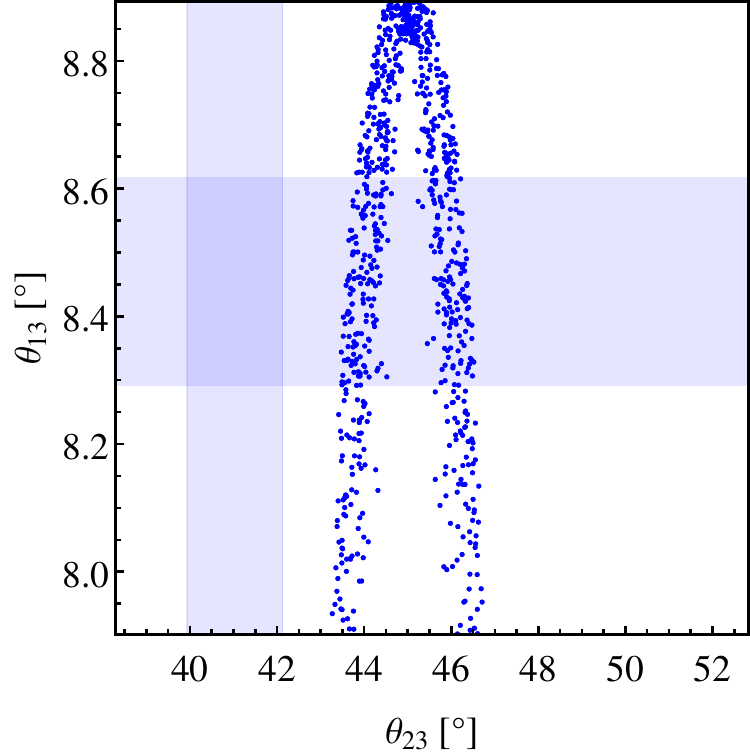}
\includegraphics[scale=0.49]{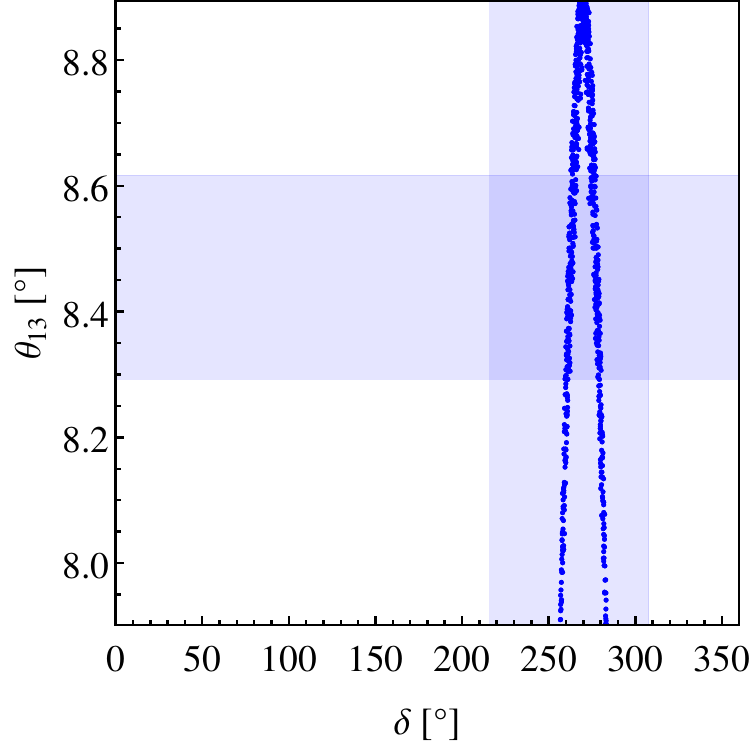}
\includegraphics[scale=0.49]{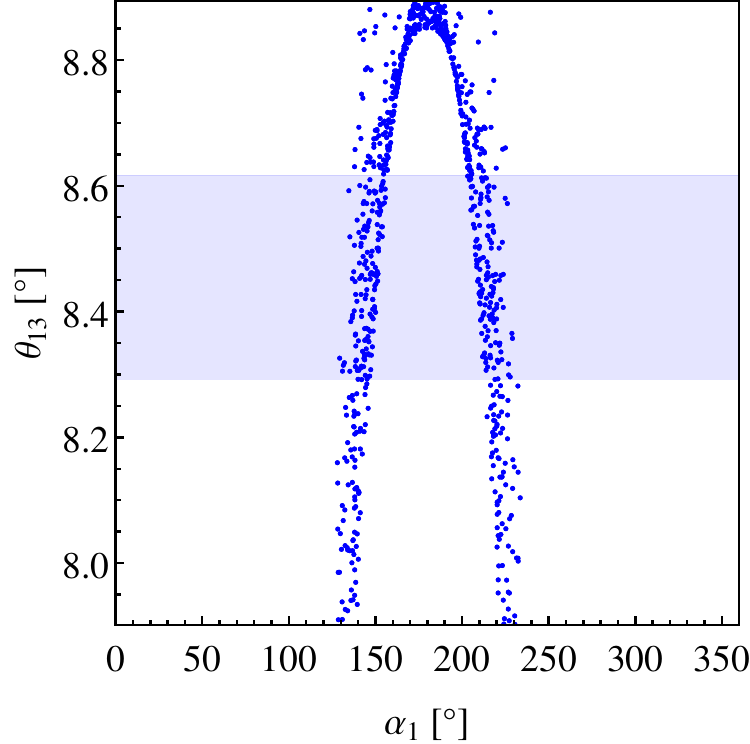}
\includegraphics[scale=0.49]{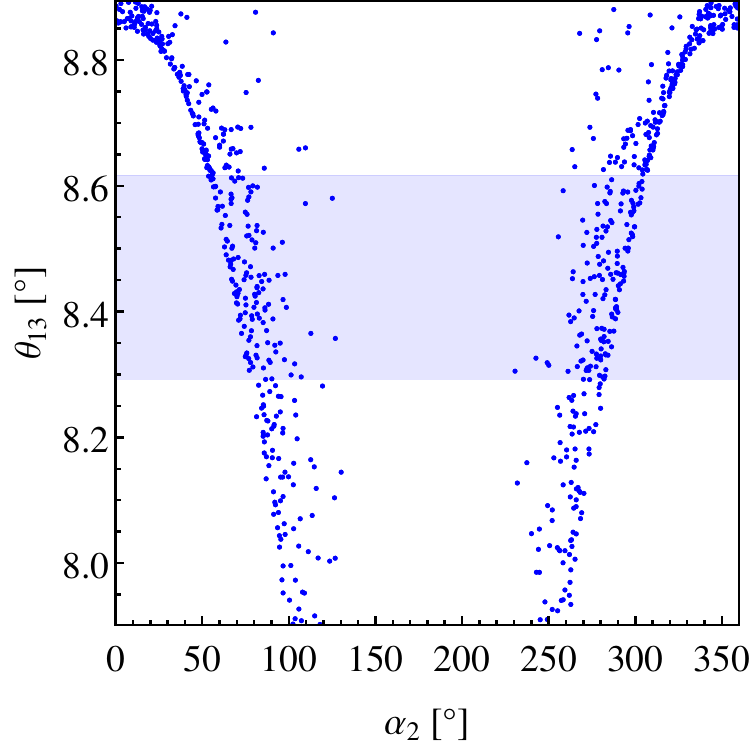}
\includegraphics[scale=0.475]{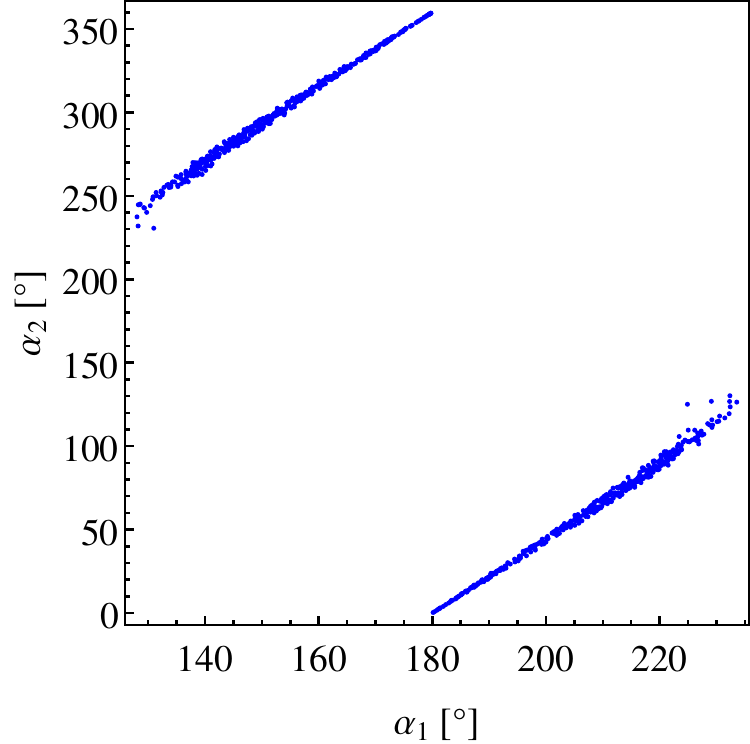}
\caption{
Numerical parameter space scan result. For observables (and also $\delta$), the plots are framed in their $3\sigma$ ranges, and the colored bands mark the $1\sigma$ ranges. The ranges are taken from Ref.~\cite{deSalas:2017kay}. Blue points given by the model are in agreement with 3$\sigma$ ranges of the low energy neutrino masses and mixing parameters. 
}
\label{fig:mixing_parameters}
\end{figure}

In Figure~\ref{fig:mixing_parameters} we show the result of the numerical parameter space scan for neutrino masses in the normal ordering. For the low energy oscillation observables (and also $\delta$), the plots are framed in their $3\sigma$ ranges, and the colored bands mark the $1\sigma$ ranges. The ranges are taken from Ref.\cite{deSalas:2017kay}. We find that there are points given by the model in agreement with $3\sigma$ ranges of the low energy neutrino masses and mixing parameters. This fact signals the $\hat{m}_{\nu_{\rm model}}$ in Eq.(\ref{eq:mmodel}) after RG running is phenomenologically viable. Given the fact that $\hat{m}_{\nu_{\rm model}}$ cannot be of exactly the SC form (which is also phenomenologically viable), we see that the effects of $\hat{m}_{\rm D}$ together with the RG running, and more importantly, the elaborately constructed $\hat{m}_{\rm R}$, render a phenomenologically viable $\hat{m}_{\nu_{\rm model}}$ in the low energy. It requires more work to disentangle these effects. We can get a sense of the RG effect by inputting the best fit values of the model parameters to the observables, and then performing a RG running down to the low energy. We find that, e.g. $\theta_{12}$ diminishes at a level of $\mathcal{O}(10^{-6})$ radian, $\theta_{13}$ and $\theta_{23}$ diminishes at a level of $\mathcal{O}(10^{-5})$ radian. This means a mild destructive effect. As we know what $\hat{m}_{\rm R}$ would give, we conclude that the $\hat{m}_{\rm D}$ effect is also mild and compensates the RG effect, at least in the case of these input. 

In the first two plots of Figure~\ref{fig:mixing_parameters}, we see deviations from the self-complementarity relation. It is understood that different sets of parameters agree with the SC relation at different level, and more importantly, the RG effects of the three mixing angles are different~\cite{Antusch:2003kp}. We prefer to scan over the parameter space to find points compatible with all the low energy constraints rather than the SC relation for obvious reasons. On one hand, it is what we build a model for; on the other hand, the reproducing of $\hat{m}_\nu$ structure guarantees the realization of the SC relation. The numerical result here clearly carries the features of the SC mixing as in $\theta_{23}$ and $\delta_{\rm CP}$.

Given the numerical result as shown in Figure~\ref{fig:mixing_parameters}, we are ready to discuss the predictions of the model. In the following we give model predictions using the $3\sigma$ low energy constraints.

The CP violating phases are predicted as 
\begin{align}
\delta_{\rm CP} \in[256.72^\circ,283.33^\circ];\\
\alpha_1 \in [128.03^\circ,233.58^\circ];\\
\alpha_2 \in [0.30^\circ,130.21^\circ] \cup [230.59^\circ,359.43^\circ].
\end{align}

The rephasing invariant in oscillation, i.e., the Jarlskog invariant~\cite{Jarlskog:1985ht}, is
\begin{align}
 J_{\rm CP}=\frac{1}{8}\cos\theta_{13}\sin2\theta_{12}\sin2\theta_{23}\sin2\theta_{13}\sin\delta \in [0.029,0.036].
\end{align} 

The sum of the neutrino masses is  
\begin{align}
\sum_j m_j \in [0.062,0.077] ~\text{eV} \;,
\end{align}
which lies safely within the cosmology limit $\sum_j m_j < 0.23 \text{ eV}$ \cite{Ade:2013zuv}.

The kinematic mass $m_{\beta}$  as measured
in the KATRIN experiment~\cite{Drexlin:2005zt} is 
\begin{align}
m_{\beta}=\sqrt{m_{1}^{2}c_{12}^{2}c_{13}^{2}+m_{2}^{2}s_{12}^{2}c_{13}^{2}+m_{3}^{2}s_{13}^{2}} \in [0.009,0.014] ~\text{eV},
\label{eq:katrin}
\end{align}
which is below the reach of $m_{\beta} > 0.2$~eV.
 
\begin{figure}
\centering
\includegraphics[scale=0.7]{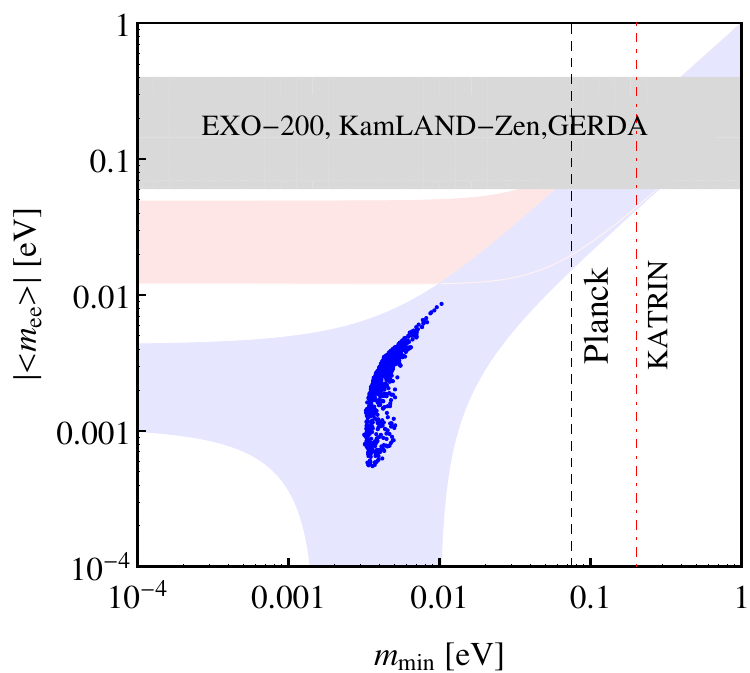}
\caption{
Prediction for the effective Majorana neutrino mass $|\langle m_{\rm ee}\rangle|$ in neutrinoless double beta decay
experiments as a function of the lightest neutrino mass $m_{\rm min}$. The blue points are given by the model respecting the 3$\sigma$ low energy constraints. The light blue (pink) region are obtained from the 3$\sigma$ ranges of the low energy
neutrino masses and mixings in case of normal (inverted) ordering. The vertical black dashed line is the Planck limit~\cite{Ade:2013zuv}, and the vertical red dot-dashed line represents the limit on $m_{\rm min}$ ($\sim0.2$ eV) obtained from KATRIN sensitivity~\cite{Drexlin:2005zt}. The light grey region is the upper limit on $|\langle m_{\rm ee}\rangle|$ given by the EXO-200~\cite{Albert:2017owj}, KamLAND-Zen~\cite{KamLAND-Zen:2016pfg}, and GERDA experiments~\cite{Agostini:2017iyd}.  
}
\label{fig:mee_plot}
\end{figure}

The effective Majorana mass in neutrinoless double beta decay is
\begin{align}
|\langle m_{\rm ee}\rangle|=|m_1c_{12}^2c_{13}^2e^{-i\alpha_1}+m_2s_{12}^2c_{13}^2e^{-i\alpha_2}+m_3s_{13}^2e^{-2i\delta}|\in [0.00001,0.010]~\text{eV}.
\end{align}
We show in Figure~\ref{fig:mee_plot} this prediction in comparison with current limits given by oscillation experiments, neutrinoless double beta decay experiments, cosmology and also the KATRIN experiment.

Notice that the above results are given when light neutrino masses are in the normal ordering. In the inverted ordering case, after performing a similar parameter space scan, we cannot find viable points given the $3\sigma$ constraints of the observables. We can get some insights into this issue by fitting the models' predictions on $\{\theta_{12}, \theta_{13}, \theta_{23}, \delta_{\rm CP}, \Delta m_{21}^2, \Delta m_{32}^2\}$ to their global fit values. In the inverted ordering case, we get a $\chi^2_{\rm min}/_{\rm NDF} \simeq 12/1$, indicating that the model is not a suitable description of the data in this case. So we conclude that this model is not viable to give realistic predictions in the inverted ordering of neutrino masses.  

\section{Summary and conclusion}\label{sec:sum}

In this paper, we construct a self-complementary mixing pattern from the self-complementarity relation and $\delta_{\rm CP}=-\frac{\pi}{2}$. A Majorana neutrino mass matrix is obtained from the perturbative SC mixing to $\mathcal{O}(\lambda^2)$ ($\lambda=\sin\theta_{13}$). Then we build an $S_4$ flavour model to reproduce the structure of the Majorana mass matrix at the high energy. After evolving down to the low energy, the model renders a phenomenologically viable effective Majorana mass matrix and gives predictions on neutrino masses and
mixing that are compatible with the oscillation experiments results in the normal ordering of light neutrino masses. For the quantities that have not been measured yet, the model gives predictions.

We argue that the SC mixing has to be perturbatively realized in a model, as it contains substructure of the $\mu-\tau$ symmetric mixing. Using the expanded SC mixing, the Majorana mass matrix is constructed order by order. Although the SC mixing has a BM mixing at the leading order, the model is not merely another BM $S_4$ model. Because the higher order terms that are carefully arranged are as important as the leading order to reveal the full structure dictated by the SC mixing. It becomes clearer when we look at the neutrino sector of the model, which contains many parameters at a first sight but the detailed structure gets most of them correlated.

It is not surprising that the model gives realistic predictions on neutrino masses and mixing parameters. The SC mixing is within the $3\sigma$ ranges of the experimental data. The charged lepton sector is separated from the neutrino sector due to the U(1) symmetry, which means no corrections from the charged lepton sector to the lepton mixing. At the same time, the symmetries of the model forbid higher order terms to mass dimension 10, so the neutrino mass matrix structure is also safe from high order corrections. There are deviations from the exact SC mixing coming from the perturbative construction to a percent level of accuracy as discussed in the end of Section~\ref{sec:sc} and the effect caused by the non-diagonal $\hat{m}_{\rm D}$ as discussed in the end of Section~\ref{sec:model}. Allowing for these deviations and taking into consideration the RG effects, we find a region in the parameter space that is compatible with all the low energy observables only in the case of normal ordering.

The model also gives predictions on the not-yet observed quantities. For example, the Dirac CP violating phase is predicted to be in the range $[256.72^\circ,283.33^\circ]$, and the Majorana phases are: $\alpha_1 \in [128.03^\circ,233.58^\circ]$, $\alpha_2 \in [0.30^\circ,130.21^\circ] \cup [230.59^\circ,359.43^\circ]$. The lightest neutrino mass is $m_1\in [0.003,0.010]$ eV. The effective Majorana neutrino mass in neutrinoless double beta decay is $|\langle m_{\rm ee}\rangle| \in [0.00001,0.010]$ eV. These quantities might be measured in future experiments.

In sum, the $S_4$ model we built are elaborately controlled at a percent level of accuracy to render the mass matrix structure dictated by the SC mixing. Also as a result of the control, there are few free parameters left in the model. A numerical study of the parameter space shows that the model gives realistic predictions on neutrino masses and mixings, and can be tested in future experiments.

\vspace{.5cm}

{\large \bf{Acknowledgement}}

The author would like to thank J. Gehrlein and M. Spinrath for sharing the code generating the $|\langle m_{\rm ee}\rangle|$
vs. $m_{\rm min}$ plot, thank J.P. Dai, I. Girardi and G.S. Li for discussions on the numerical method, and thank Prof. X.G. He for his hospitality in SJTU where the work was done. This work is supported in part by the Shanghai Laboratory for Particle
Physics and Cosmology under Grant No. 11DZ2260700.
 
\appendix 

\section{Correlation of the model parameters}\label{apd:para}
The parameters shown in the neutrino sector of the model in Section~\ref{sec:model} can be reduced significantly. There are two kinds of parameters redundancy: First, many parameters are correlated when we restrict the heavy Majorana neutrino mass matrix to be in the form of $\hat{m}_\nu$ to order $\lambda^2$ as shown in Section~\ref{sec:sc}; Second, for undetermined parameters that are multiplied together exclusively, they are effectively one independent parameter. We use a parameter redefinition to get rid of the latter kind of redundancy. In the following we show the details on how this is achieved.

By our construction in Section~\ref{sec:model}, $\hat{m}_{\rm R}$ is in the same form as the $\hat{m}_\nu$ gets from the SC mixing. We express it as
\begin{align}
\hat{m}_\nu|_{\text{to}~\mathcal{O}(\lambda^2)}=t\hat{m}_{\rm R}|_{\text{to}~\rm NNLO},
\end{align}
where t includes potential scaling difference. We impose such a ``bizarre" relation so that after seesaw the construction will not be ruined. It is not the first time such a relation shows up. In discrete flavour symmetry models, it is more often in the form that the heavy and light neutrino mass matrices can be diagonalized by the same mixing matrix. 

We further require that this equation holds for each order. In this way, we have
\begin{align}
t\frac{y_{13}}{\Lambda}v_{\phi_1}=\frac{m_1-m_2}{2\sqrt{2}};\\
t\frac{y_{12}}{\Lambda}v_{\psi_1}=\frac{1}{2\sqrt{3}}(m_1+m_2+2m_3);\\
t\frac{y_{21}}{\Lambda^2}v_{\phi_{21}}^2=-\lambda a;\\
t\frac{y_{22}}{\Lambda^2}v_{\phi_{22}}^2=i\lambda a;\\
t\frac{y_{23}}{\Lambda^2}v_{\phi_{23}}^2=-\frac{1}{2}i\lambda b;\\
t\frac{y_{31}}{\Lambda^3}v_{\psi_3}^2v_{\xi_3}=\frac{2}{3}\lambda^2c;\\
t\frac{y_{32}}{\Lambda^3}v_{\phi_{31}}^3=-\frac{1}{2}\lambda^2d;\\
t\frac{y_{33}}{\Lambda^3}v_{\phi_{32}}^3=-\frac{1}{2}i\lambda^2e,
\end{align} 
where a,b,c,d,e are defined in Eq.~(\ref{eq:ab}) and Eq.~(\ref{eq:cde}). In the above expressions, all the quantities except t, $\lambda$, $\Lambda$ are complex, and the imaginary unit ``i" should cause no confusion. From these expressions, we get the ratios
\begin{align}
\frac{y_{21}(v_{\phi_{21}}/\Lambda)^2}{y_{13}(v_{\phi_1}/\Lambda)}=-\sqrt{2}\lambda\equiv\frac{\tilde{v}_{\phi_{21}}^2}{\tilde{v}_{\phi_1}};\label{eq:v21tv1}\\
\frac{y_{22}(v_{\phi_{22}}/\Lambda)^2}{y_{13}(v_{\phi_1}/\Lambda)}=i\sqrt{2}\lambda\equiv\frac{\tilde{v}_{\phi_{22}}^2}{\tilde{v}_{\phi_1}};\\
\frac{y_{32}(v_{\phi_{31}}/\Lambda)^3}{y_{13}(v_{\phi_1}/\Lambda)}=-\frac{5}{4}\lambda^2\equiv\frac{\tilde{v}_{\phi_{31}}^3}{\tilde{v}_{\phi_1}};\\
\frac{y_{33}(v_{\phi_{32}}/\Lambda)^3}{y_{13}(v_{\phi_1}/\Lambda)}=i\lambda^2\equiv\frac{\tilde{v}_{\phi_{32}}^3}{\tilde{v}_{\phi_1}};\\
\frac{y_{23}(v_{\phi_{23}}/\Lambda)^2}{y_{12}(v_{\psi_1}/\Lambda)}=\frac{\sqrt{6}}{4}i\lambda\equiv\frac{\tilde{v}_{\phi_{23}}^2}{\tilde{v}_{\psi_1}};\\
\frac{y_{31}(v_{\psi_3}^2v_{\xi_3}/\Lambda^3)}{y_{12}(v_{\psi_1}/\Lambda)}=-\frac{\sqrt{3}}{3}\lambda^2\equiv\frac{\tilde{v}_{\psi_3}^2\tilde{v}_{\xi_3}}{\tilde{v}_{\psi_1}}\label{eq:v33tv1},
\end{align}
where in the last step of each equation we redefined the parameters by absorbing the scale factor $\Lambda$ and the dimension $1$ coupling $y_{ij}$ to the vevs of the flavons. We also define $\frac{y_{11}}{\Lambda}v_{\xi_1}\equiv\tilde{v}_{\xi_1}$. In this way, all the structural information will be contained in the redefined vevs $\tilde{v}$, and there are three independent $\tilde{v}$: $\tilde{v}_{\xi_1}$, $\tilde{v}_{\psi_1}$ and $\tilde{v}_{\phi_1}$. Notice that these $\tilde{v}$ are all complex. To reproduce the relation $(\hat{m}_0)_{23}=(\hat{m}_0)_{11}-(\hat{m}_0)_{22}$, the phase of $\tilde{v}_{\xi_1}$ should be the same as the phase of $\tilde{v}_{\psi_2}$. We choose their common phase as $\gamma$, and the phase of $\tilde{v}_{\phi_1}$ as $\rho$. To avoid the notation complexity, after the definition of the phases, we take $\tilde{v}$ as real. So we conclude that we are left with five real parameters: $\tilde{v}_{\xi_1}$, $\tilde{v}_{\psi_1}$ and their phase $\gamma$; $\tilde{v}_{\phi_1}$ and its phase $\rho$.

\section{Vacuum alignment}\label{sec:vev}
In this section we exhibit how the flavon vevs used in Section~\ref{sec:model} can be obtained from a renormalisable scalar potential. We first look at all the invariant terms under the model's full symmetries:

\begin{align}
&(\xi^\dagger \xi)_1,\quad (\psi^{a\dagger} \psi^a)_1,\quad(\phi^{a\dagger} \phi^b)_1,\nonumber\\
&(\xi^\dagger\xi\xi^\dagger\xi)_1,\quad (\xi^\dagger\xi)_1(\psi^{a\dagger}\psi^a)_1,\quad(\xi^\dagger\xi)_1(\phi^{a\dagger}\phi^b)_1,\quad (\psi^{a\dagger}\psi^a)_1(\psi^{b\dagger} \psi^b)_1,\quad(\psi^{a\dagger} \psi^a)_2(\psi^{b\dagger} \psi^b)_2,\nonumber\\
&(\psi^{a\dagger} \psi^a)_1(\phi^{b\dagger} \phi^c)_1,\quad (\psi^{a\dagger} \psi^a)_2(\phi^{b\dagger} \phi^c)_2,\nonumber\\
&(\phi^{a\dagger} \phi^b)_1(\phi^{a\dagger} \phi^c)_1,\quad (\phi^{a\dagger} \phi^b)_2(\phi^{a\dagger} \phi^c)_2,\quad (\phi^{a\dagger} \phi^b)_3(\phi^{a\dagger} \phi^c)_3,\quad (\phi^{a\dagger} \phi^b)_{3^\prime}(\phi^{a\dagger} \phi^c)_{3^\prime}. \nonumber
\end{align}

The $S_4$ singlet, doublet and triplet are denoted as $\xi, \psi, \phi$ separately, and the $S_4$ contractions are shown in the number outside the parentheses. ``a, b, ..." are used to distinguish different flavons and ``a=b=..." are allowed whenever they show up.\footnote{The quartic term $(\phi^{a\dagger} \phi^b)_{3^\prime}(\phi^{a\dagger} \phi^c)_{3^\prime}$ only has a non-zero value for the case $a\neq b\neq c$.} Notice that the singlet and doublet only contract with itself first, while triplet can contract with a different triplet first. It only happens for the flavons $\phi_{2i}, i=1,2,3$ due to their $\mathbb{Z}_2$ charges, thus only three different flavons can appear in the quartic triplet terms. 

The invariant terms are all in a form, e.g., $\psi^\dagger\psi$, to guarantee the invariance under the $U(1)$ and 
$\mathbb{Z}_n$ symmetries. These symmetries forbid terms with mass dimension 3. Also with this combination, the overall phases disappear.\footnote{The magnitudes and phases of flavons are discussed in Appendix~\ref{apd:para}.} We will see in the following the expressions with real components of each flavons.

The most general potential contains the invariant terms for each flavon and all their possible mixings. In our case, we can have at most three flavons mixing, which is only valid for the flavons $\phi_{2i}, i=1,2,3$. In the following, we construct scalar potentials as subsets of the most general potential, only for illustrative purpose. 

First consider a potential for a singlet $\xi$ and two doublets $\psi^a=(\psi_1^a,\psi_2^a), \psi^b=(\psi_1^b,\psi_2^b)$
\begin{align}
V&\supset  c_1 \left((\psi_1^a)^2+(\psi_2^a)^2\right)+ c_2 \left((\psi_1^b)^2+(\psi_2^b)^2\right)\nonumber\\
&+ c_3 \left((\psi_1^a)^2+(\psi_2^a)^2\right)^2 +c_4\left(4(\psi_1^a)^2(\psi_2^a)^2+\left(-(\psi_1^a)^2+(\psi_2^a)^2\right)^2\right)\nonumber\\
&+c_5 \left((\psi_1^b)^2+(\psi_2^b)^2\right)^2 +c_6\left(4(\psi_1^b)^2(\psi_2^b)^2+\left(-(\psi_1^b)^2+(\psi_2^b)^2\right)^2\right)\nonumber\\
&+ c_7\left((\psi_1^a)^2+(\psi_2^a)^2\right)\left((\psi_1^b)^2+(\psi_2^b)^2\right),
\end{align}
after imposing the first and second derivative tests, we find that the potential has a local minimum at 
$\psi^a\sim (3,1),\quad \psi^b\sim (-\sqrt{3},1)$, for
\begin{align}
c_1+20c_3+20c_4+4c_7=0,\\
c_2+8c_5+8c_6+10c_7=0,\\
c_5+c_6>0,\\
c_3+c_4>0.
\end{align}

We could also write a scalar potential of the form
\begin{align}
V\supset d_1 \left((\phi_1^a)^2+2\phi_2^a\phi_3^a\right)+ d_2\left((\phi_1^a)^2+2\phi_2^a\phi_3^a\right)^2.
\end{align}
It reaches a minimum at $\phi^a\sim(1,1,1)$ for $d_1=-6d_2, d_2>0$.

Consider now a scalar potential like
\begin{align}
V&\supset d_1 \left((\phi_1^a)^2+2\phi_2^a\phi_3^a\right)+d_2 \left((\phi_1^b)^2+2\phi_2^b\phi_3^b\right)\nonumber\\
&+d_3 \left((\phi_1^a)^2+2\phi_2^a\phi_3^a\right)\left((\phi_1^b)^2+2\phi_2^b\phi_3^b\right)\nonumber\\ &+d_4\left(\left((\phi_1^a)^2-2\phi_2^a\phi_3^a\right)\left((\phi_1^b)^2-2\phi_2^b\phi_3^b\right)+\frac{3}{4}\left((\phi_2^a)^2+(\phi_3^a)^2\right)\left((\phi_2^b)^2+(\phi_3^b)^2\right)\right)\nonumber\\
&+d_5 \left(-8(\phi_1^a)^2\phi_2^a\phi_3^a+\left(-(\phi_2^a)^2+(\phi_3^a)^2\right)^2\right)+d_6 \left(-8(\phi_1^b)^2\phi_2^b\phi_3^b+\left(-(\phi_2^b)^2+(\phi_3^b)^2\right)^2\right).
\end{align}
It has a minimum at $\phi^a\sim(0,1,1), \phi^b\sim(0,-1,1)$ for
\begin{align}
d_1=-d_2=2d_3-d_4,\\
3d_4+8d_5>0, \\
3d_4+8d_6>0,\\
d_5<0, \\
d_6>0.
\end{align}

For scalar potential like
\begin{align}
V&\supset
d_1\left(((\phi_1^a)^2-\phi_2^a\phi_3^a)^2+\frac{3}{4}((\phi_2^a)^2+(\phi_3^a)^2)^2\right)+d_2\left(-8(\phi_1^a)^2\phi_2^a\phi_3^a+(-(\phi_2^a)^2+(\phi_3^a)^2)^2\right)\nonumber\\
&+d_3\left(((\phi_1^b)^2-\phi_2^b\phi_3^b)^2+\frac{3}{4}((\phi_2^b)^2+(\phi_3^b)^2)^2\right)+d_4\left(-8(\phi_1^b)^2\phi_2^b\phi_3^b+(-(\phi_2^b)^2+(\phi_3^b)^2)^2\right)\nonumber\\
&+d_5\left(((\phi_1^a)^2-\phi_2^a\phi_3^a)((\phi_1^b)^2-\phi_2^b\phi_3^b)+\frac{3}{4}((\phi_2^a)^2+(\phi_3^a)^2)((\phi_2^b)^2+(\phi_3^b)^2)\right)\nonumber\\
&+d_6\left(-4\phi_1^a\phi_3^a\phi_1^b\phi_2^b-4\phi_1^a\phi_2^a\phi_1^b\phi_3^b+(-(\phi_2^a)^2+(\phi_3^a)^2)(-(\phi_2^b)^2+(\phi_3^b)^2)\right),
\end{align}
we get a minimum at $\phi^a\sim(0,0,1), \phi^b\sim(0,1,0)$ for
\begin{align}
3d_1+4d_2+\frac{3}{2}d_5-2d_6=0,\\
3d_3+4d_4+\frac{3}{2}d_5-2d_6=0,\\
3d_1+4d_2>0,\\
3d_3+4d_4>0,\\
d_1-4d_2+2d_6>0,\\
d_3-4d_4+2d_6>0,\\
d_1>0,\\
d_3>0.
\end{align}

Now we have all the flavons alignments used in the model. In our construction there are some invariant terms omitted. The omitted terms should have good reasons for not showing up, which is highly nontrivial in such kind of model building. The possible ways out could be extra symmetries or localizing in different branes when using an extra dimension set~\cite{Merle:2013gea}. 

It may be useful to remind the reader that throughout this section, we discuss only the alignments. The magnitudes and the overall phases, which are combined in the complex quantities, e.g., $v_\phi$, are discussed in Appendix~\ref{apd:para}. They are related to $\lambda$ and the scale $\Lambda$ through Eq.(\ref{eq:v21tv1}) to Eq.(\ref{eq:v33tv1}).

\end{document}